\documentstyle[12pt,epsf]{article}
\catcode`\@=11
\def\beq{\begin{equation}}
\def\eeq{\end{equation}}
\def\beqa{\begin{eqnarray}}
\def\eeqa{\end{eqnarray}}
\def\rd{{\mathrm d}}

\def\intd4x{\int{\rd}^4x}

\def\m32{{m_{3/2}}}

\begin{document}

\begin{titlepage}
\begin{flushright}
FERMILAB-Pub-94/075-A\\
DFPD 94/TH/21
\hep-ph /9404228
\end{flushright}

\begin {center}
{\Huge Inclusive particle photoproduction to next-to-leading order}

\end{center}
\vspace{2cm}
\begin{center}
{\bf M. Greco$^{a}$, S. Rolli$^{b}$, A. Vicini$^{c}$}\\
\vspace{1cm}
{$^a$\it Dipartimento di Fisica, Universit\`a dell'Aquila, Italy}\\
{\it INFN, Laboratori Nazionali di Frascati, Frascati, Italy}\\
\vspace{0.5cm}
{$^b$\it Dipartimento di Fisica Nucleare e Teorica,,Universit\`a di Pavia,
Italy}\\
{\it INFN, Sezione di Pavia, Pavia, Italy and}\\
{\it NASA/Fermilab Astrophysics Center, FERMILAB, Batavia, Il 60510}\\
\vspace{0.5cm}
{$^c$\it Dipartimento di Fisica, Univerist\`a di Padova, Italy}\\
{\it INFN, Sezione di Padova, Padova, Italy}\\
\end{center}
\vspace{2cm}
\begin{center}
{\bf Abstract}
\end{center}
We study the inclusive
photoproduction of neutral and charged pions and $\eta$ at HERA, via
the resolved photon mechanism, in QCD to next-to-leading order.
We present various distributions of phenomenological interest and study
the theoretical uncertainties due to the mass scales, and to photon and proton
sets of structure functions.
A new set of fragmentation functions for charged pions is also presented.

\vspace{0.5cm}
\centerline{ {\it Submitted to Zeitschrift f\"ur Physik C}}

\vfill
\end{titlepage}

Inclusive production of high $p_t$ particles and jets at HERA plays an
important role in testing QCD, providing a detailed source of
information on the hadron-like structure of the photon.

For this purpose leading order (hereafter denoted as LO )
perturbative QCD
predictions - based on
evalations of partonic cross sections at tree level and evolution of
structure and fragmentation functions at one loop level--~
are not accurate enough, being plagued by the
usual theoretical uncertainties associated to the large scale
dependence of $O(\alpha_{em}\alpha_S)$ terms.
A consistent calculation at next-to-leading order
(hereafter denoted as NLO) needs two loop evolved structure and fragmentation
functions and a NLO evaluation of
parton-parton subprocesses.

As well known,two mechanisms contribute to the inclusive photoproduction
of particles or jets at high energies: the photon can interact directly with
the
partons originating from the proton (direct process), or
via its quark and gluon content (resolved process).

Previous theoretical analyses have considered both direct photoproduction to
NLO, Aurenche et al. \cite{au}, and resolved photoproduction, Borzumati et al.
\cite{franc}, the latter having used
the NLO corrections to all contributing parton-parton scattering
processes of Aversa et al.\cite{aversa}, and LO fragmentation functions for
the final
hadron. Those results show the dominance of the resolved component at low
$p_t$
($p_t <$  10 Gev), which is the region firstly explored at HERA, the role
played
by the direct contributions being shifted at higher $p_t$. The separation of
the
cross section in two components induces an artificial dependence on the photon
factorization mass scale $M_{\gamma}$, which should cancel when the two terms
are added up.

Indeed this mechanism has been explicitly shown \cite{vik} to apply in the
inclusive
photoproduction of jets, which has been recently studied to NLO accuracy.
Furthermore, the results of ref.~\cite{vik} also differ from previous NLO
studies of the
resolved component, the various analyses giving predictions not in full
agreement
even at the LO level.

Motivated by these results, we consider in this paper the photoproduction of
single hadrons in electron-proton collisions at HERA energies, based on the
recent NLO fragmentation functions of ref ~\cite{noi}, limiting ourselves to
the
study of the resolved component only. A full NLO analysis including the direct
term will be given elsewhere \cite{work}.

In particular
we present a detailed quantitative evaluation of $\pi^0$, $\pi^{\pm}$
and $\eta$
photoproduction at HERA at moderate $p_t$, using the hard
scattering cross sections of ref.~\cite{aversa}, and two loop
structure and fragmentation functions. While the $\pi^0$ and $\eta$
fragmentation functions have been discussed earlier \cite{noi,noieta}, a new
set of fragmentation functions for charged pions is presented here as well.

We give now the relevant formulae for the cross sections.
The inclusive cross section for $ep\to h + X$ in an
improved next-to-leading-order approximation is:
\beq
E_{h}{{ d^3\sigma(ep\to h +X)}\over{d^3p_{h}}}=
\int_{x_{min}}^{1}{} dxf_{\gamma/e}(x)\hat{E_h}{{
d^3\hat{\sigma}(\gamma p\to h +X)}\over{d^3\hat p_{h}}}(x)
\eeq
where $x_{min}$ is given in terms of the transverse momentum $p_t$
and of the center-of-mass pseudorapidity $\eta_{cm}$ of the
produced hadron as:
\beq
x_{min} = {{p_t e^{\eta_{cm}}}\over{\sqrt{s} - p_t e^{-\eta_{cm}}}}
\eeq
The rapidity  $\eta_{lab}$ measured in the laboratory frame  is related to
$\eta_{cm}$ as:
\beq
\eta_{lab}=\eta_{cm}-{1\over 2}\ln {{E_p}\over{E}}
\eeq
where $E$ and $E_p$ are the energies of the electron and the proton
respectively
($E= 27$ GeV and
$E_p=820$ GeV, for the present HERA conditions).

The distribution in the longitudinal momentum fraction $y$ of the
outgoing photon has in the NLO approximation the following form \cite{frix}:
\[
 f_\gamma^{(e)}(y)~=~\frac{\alpha_{em}}{2\pi}\left\{
2(1-y)
\left[\frac{m_e^2y}{E^2(1-y)^2\theta_c^2 + m_e^2y^2}-\frac{1}{y}\right]\right.
{}~~~~~~~~~~~~~~\]
\beq
\left.~~~~~~~~~~~~~~~~~~+\frac{1+(1-y)^2}{y}
\log\frac{E^2(1-y)^2\theta_c^2 +m_e^2 y^2}{m_e^2 y^2}
+{\cal O}(\theta_c^2,m_e^2/E^2)\right\},
\eeq
where
$\theta_c$
is the maximum value of the electron scattering angle.

Finally the $\gamma p$ inclusive cross section is given by:
\[
\label{eq:cs}
 E{{d\sigma^{\gamma p}}\over{d^3P}}=
{1\over{\pi
S}}\sum_{i,j,l}\int_0^1\int_0^1\int_0^1dx_1~dx_2~{{dx_3}\over{x_3^2}}
F_i^p(x_1,M^2_p)F_j^{\gamma}(x_2,M^2_{\gamma})D_{l}^H(x_3, M_f^2)
\times~~~~~~~~~~~~
\]
\begin{equation}
{}~~~~\times\left({{\alpha_S(\mu^2)}\over{2\pi}}\right)^2
\left[{1\over v}\sigma_{ijl}^0(s,v)\delta(1-w)
+{{\alpha_S(\mu^2)}\over{2\pi}}K_{ijl}(s,v,w;M_p^2,M_{\gamma}^2,\mu^2,M_f^2)
\right]
\end{equation}
where $s,v$ and $w$ are the partonic variables
$s=x_1x_2S,~v={{x_2-1+V}\over{x_2}},~w={{x_2VW}\over{x_1(x_2-1+V)}}$ and
$V=1+{T\over S},~W={{-U}\over{T+S}}$, with $S,T,U$
the hadronic Mandelstam variables.
$\sigma_{ijl}^0$ are the partonic Born cross sections $O(\alpha_S^2)$, while
$K_{ijl}$ are the finite higher order corrections $O(\alpha_S^3)$
\cite{aversa},with  $i,~j,~l$ running on all kinds of partons.
In principle we have kept distinct all
factorization mass scales of the structure and fragmentation functions.
As usual, the photon structure functions are expressed in terms of the hadronic
and the pointlike contributions as
$F^{\gamma}(x,Q^2)=F^{\gamma}_{had}(x,Q^2)+F^{\gamma}_{point}(x,Q^2)$,
and obey the appropriate evolution equation with the inhomogeneous
term related to $F^{\gamma}_{point}$.

As already stated above a consistent calculation to next-to-leading order
needs two-loop evolved structure and fragmentation
functions
and a NLO evaluation of
parton-parton subprocesses.
In the partonic cross sections to one loop \cite{aversa},
calculated from the squared matrix elements
$O(\alpha_S^3)$ of Ellis et Sexton \cite{ellis}, the initial
state collinear divergences
have been factorised and absorbed into the dressed structure functions in
the $\overline{MS}$ scheme.
Coherently with this choice, we have used for the proton structure functions
 set B1 of Morfin \& Tung,
\cite{mt}, set  MRS S0 of Martins Roberts \& Stirling \cite{mrs},
and set GRV HO of Gl\"uck, Reya \& Vogt \cite{grvpro}
 and three different NLO parametrisations of the
photon structure functions, namely
the set of Aurenche et al. \cite{aurenche}  (set I), that of
Gl\"uck, Reya and Vogt
(set II) \cite{grv} and that of
Gordon and Storrow,
 (set III)\cite{gs}.
Sets I and II have been also used in the previous analysis of Borzumati et al.
\cite{franc}.

We have used the improved expression (4) for the Weiszaecher- Williams
photon density in the electron \cite{frix}. When comparing our results with
those obtained with the usual leading order formula (e.g. see eq.1 in
ref. \cite{baer}),
and we found a negative correction which is no larger than 5\%.

Fragmentation functions will be also considerd to NLO accuracy. For the
$\pi^0$
case, various consistent parametrizations have been discussed in
ref.~\cite{noi},
using different methods and initial conditions. All of them have been
successfully compared with the current experimental data in $e^+ e^-$ and
$p\bar p$
collisions at various energies. In the following we will use only one set
of them, based on the MonteCarlo simulator HERWIG \cite{herwig,herwig2}
, which is used to fix the
initial conditions at the fragmentation scale $M_0$ = 30 GeV. The same method
has also been applied in ref.\cite{noieta} to inclusive $\eta$ production, and
indeed the predicted
$\eta/\pi^0$ ratio has been found to agree with the present experimental
information at ISR \cite{isr} and from $e^+ e^-$ and $p\bar p$ colliders.
Recent
fixed target experiments also agree \cite{e706} with the predictions of ref.
\cite{noieta}. We are therefore quite confident with the reliabilty of the
method used
and consequently follow the same tecnique to
obtain a new set of fragmentation functions
NLO for $\pi^{\pm}$.  The functions are parametrized as:
\beq
D^{\pm}_i(z, M_f^2)=N_iz^{\alpha_i}(1-z)^{\beta_i}
\eeq
where $i$ runs over $(u,~d,~s,~c,~b,~g)$, and $M_0$ = 30 GeV.
The coefficients are given
in Table I.
We remark that we are considering the inclusive production of ($\pi^+$ +
$\pi^-$). We have used a new improved version of HERWIG, which also
gives us a new $\pi^0$ fragmentation set which is slightly different from that
given in ref.~\cite{noi}, and actually improves the comparison with
$e^+ e^-$ data at
high $z$ ($z >~0.7$). However, to the aim of the present work, the two sets of
$\pi^0$ fragmentation functions based on the old and new versions of HERWIG
are equivalent. For the sake of completeness we also give in Table II the
new parametrization of $\pi^0$ quark fragmentation functions at $M_0$ = 30 Gev.

We present now various numerical results for the three sets of photon structure
functions, studying in particular the uncertainties of the theoretical
predictions. Let us consider $\pi^0$ photoproduction first.

The dependence of the  cross section on the various mass
scales involved in (5) is shown in figs.~1. As expected, the dependence
is very strong at the Born level, as shown in fig.1a for
$p_T=5~GeV$,
for $\eta_{lab}=-2$. The introduction of higher orders reduces the effect,
although the dependence on the photon factorization scale only is still
important (figs.1b-1c), unlike to what is
observed in the case of hadron-hadron collisions \cite{aversa,end}.
This behaviour has been also observed in the photoproduction of jets at
HERA \cite{gs,krasal,vik}
and the photon mass scale dependence is reduced when the direct and resolved
terms are both considered \cite{vik}.

The above effect is similar for the three sets of photon structure functions.

In order to show the general $p_T$ behaviour of the cross section,
${{d\sigma}\over{d\eta dp_T}}$ is plotted in figs. 2 and fig.3 for
different values of $\eta_{lab}$,
$\mu=M_p=M_{\gamma}=M_f=p_T$, and for the three sets.

In figs.4 we present the $\eta_{lab}$ distribution for fixed $p_t=5$ GeV.
In fig. 4a the contributions is shown  by the various partonic subprocesses,
while the differential cross-sections
${{d\sigma}\over{d\eta dp_T}}$ for the three sets of photon structure
functions are compared in fig. 4b. As in the case of inclusive jet
photoproduction \cite{vik} the contribution from the gluon content of
the photon is too tiny to be observed in most of the phase space available.
On the contrary, the gluon contribution from the proton structure function
plays  a relevant role, and is essentially independent from the photon
and proton structure functions, as also shown in fig.5,
where the $\eta_{lab}$ distribution of the subprocess
$q(\gamma) g(p)\to \pi^0 + X$ for three different parametrization NLO of
proton structure functions. We have defined Set A for MT B1
\cite{mt}, Set B for MRS S0 \cite{mrs} and Set C for GRV HO
\cite{grvpro}.

We finally show the cross section integrated over different ranges of
$\eta_{lab}$ in fig. 6, for the Set I of photon structure functions,
which is of immediate phenomenological interest for HERA experiments.

Concerning the photoproduction of $\eta$ and charged pions,
we present in Figs. 7 the $p_T$ distribution for different values of
$\eta_{lab}$, in Figs. 8 the $\eta_{lab}$
distributions for $p_t= 5$ GeV and for different subprocesses, and in
Figs. 9 the distribution in $p_t$ integrated in $\eta_{lab}$.
The dependence on the photon structure functions is similar to what
found for the $\pi^0$ case.
Finally we show in Table III our prediction for the ratio $\eta/\pi^0$,
and in Fig. 10 the dependence of the cross section on the proton structure
functions.

Our results agree, within a factor of two, with the previous analysis
\cite{franc}
of inclusive neutral pions production.

To conclude,
a next-to-leading order calculation of inclusive neutral and
charged pions and $\eta$
production
in electron-proton collisions has been presented, particularly via the
resolved photon mechanism.

We have studied the effects of the theoretical uncertainties
related to the photon structure functions, as well as
the dependence from the various mass scales, which is still significative in
the
considered $p_t$ range. The inclusion of the direct component should
make this effect weaker.
We have also presented a new parametrization of $\pi^{\pm}$ fragmentation
functions.

Finally we stress that the gluon content of the proton can be
accurately disentangled via the photoproduction of single particles at HERA.
\vskip1truecm
When completing this work the paper "Inclusive particle production at Hera:
resolved and direct quasi-real photon contribution in next-to-leading
order QCD", by B.A.Kniehl and G.Kramer \cite{kk}, has appeared, where a
similar analysis has been carried out, including the direct photon
contribution and using LO fragmentation functions.

\vskip1truecm
We thank Giovanni Abbiendi for providing us the new version HERWIG57,
when it was also in a preliminary phase.

This work was supported in part by the DOE and by the NASA (NAGW-2831) at
Fermilab.

\vfill\eject

{\bf Table Captions}
\vspace{24pt}

\begin{itemize}
\item{Table I:} parameters of the $\pi^{\pm}$ fragmentation functions
at $M_0=30$ GeV (see eq.6).
\item{Table II:} parameters of the quarks fragmentation functions
into $\pi^0$ at $M_0=30$ GeV (see text).
\item{Table III:} Ratio $R=\eta/\pi^0$ for different values of $p_t$.
\end{itemize}
\vspace{0.5cm}

{\bf Figure Captions}
\begin{itemize}

\item{Figs. 1:} $\pi^0$ production.
Dependence of ${{d\sigma^{(ep)}}\over{d\eta dp_t}}$ on
the renormalization, factorization and fragmentation mass scales for
$\eta_{lab}=-2$, a) $\mu=M_p=M_{\gamma}=M_f=\xi p_t$
for $p_t=5$ GeV; b) $M_{\gamma}=\xi_{\gamma} p_t$ and
$M_p=M_{f}=\mu=p_t$  and for $p_t=5$ GeV; c)
same as b) but for $p_t=15$ GeV.
Continued line: Born, dashed line: NLO.

\item{Figs. 2:} $\pi^0$ production.
$p_t$ distributions of ${{d\sigma^{(ep)}}\over{d\eta dp_t}}$
for different values of
$\eta_{lab}$. a) Set I; b) Set II; c) Set III.

\item{Fig. 3:} $\pi^0$ production.
$p_t$ distributions of ${{d\sigma^{(ep)}}\over{d\eta dp_t}}$
for $\eta_{lab}=-2$ for the three
sets of photon structure functions.

\item{Fig. 4:} $\pi^0$ production.
$\eta_{lab}$ distributions of
${{d\sigma^{(ep)}}\over{d\eta dp_t}}$, for $p_t=5$ GeV,
for the partonic
subprocesses: a) Set I; b) comparison
of the total cross sections ( sum of all the subprocesses) for the
three sets.

\item{Fig. 5:} $\pi^0$ production.
$\eta_{lab}$ distributions of
${{d\sigma^{(ep)}}\over{d\eta dp_t}}$, for $p_t=5$ GeV, for the subprocess
$q(\gamma) g(p)\to \pi^0 + X$ for different sets of proton structure
functions.

\item{Fig. 6:} $\pi^0$ production.
$p_t$ distributions of ${{d\sigma^{(ep)}}\over{dp_t}}$
for different ranges of integrations over $\eta_{lab}$.

\item{Fig. 7:} $p_t$ distributions of ${{d\sigma^{(ep)}}\over{d\eta dp_t}}$
for $\eta_{lab}=-2$: a) $ep\to \eta+ X$; b) $ep\to \pi^{\pm}+ X$.

\item{Fig. 8:} $\eta_{lab}$ distributions of
${{d\sigma^{(ep)}}\over{d\eta dp_t}}$
for $p_t=5$ GeV: a) $ep\to \eta+ X$; b) $ep\to \pi^{\pm}+ X$.

\item{Fig. 9:} $p_t$ distributions of ${{d\sigma^{(ep)}}\over{dp_t}}$
for different ranges of integrations over $\eta_{lab}$:
a) $ep\to \eta+ X$; b) $ep\to \pi^{\pm}+ X$.

\item{Fig.10:} $\pi^0$ production. $p_t$ distributions of
${{d\sigma^{(ep)}}\over{d\eta dp_t}}$ for $\eta_{lab}=-2$ and set I for
the photon structure functions; sets A, B and C are used for the proton
structure functions (see text).

\end{itemize}
\vfill\eject
$$
\begin{tabular}{|c|c|c|c|c|}
\hline
\hline
\multicolumn{1}{|c|}{$Parton$} &
\multicolumn{1}{c|}{$\alpha$} & \multicolumn{1}{c|}{$\beta$}
&\multicolumn{1}{c|}{$N_i$}
&\multicolumn{1}{c|}{$<n_i>$}\\
\hline
u & $-1.14 \pm 0.01$ & $2.43 \pm 0.03$ & 1.09 & 5.15  \\ \hline
d & $-1.16 \pm 0.01$ & $2.26 \pm 0.04$ & 1.02 & 5.14  \\ \hline
s & $-0.90 \pm 0.01$ & $5.76 \pm 0.06$ & 2.93 & 4.91  \\ \hline
c & $-0.80 \pm 0.01$ & $7.52 \pm 0.09$ & 5.56 & 5.98  \\ \hline
b & $-1.14 \pm 0.01$ & $8.40 \pm 0.06$ & 2.37 & 7.41  \\ \hline
g & $-0.51 \pm 0.01$ & $5.02 \pm 0.09$ & 11.13& 6.27  \\ \hline
\end{tabular}
$$
\begin{center}
{\bf Table I}\\
\end{center}

$$
\begin{tabular}{|c|c|c|c|c|}
\hline
\hline
\multicolumn{1}{|c|}{$Parton$} &
\multicolumn{1}{c|}{$\alpha$} & \multicolumn{1}{c|}{$\beta$}
&\multicolumn{1}{c|}{$N_i$}
&\multicolumn{1}{c|}{$<n_i>$}\\
\hline
u & $-1.18 \pm 0.01$ & $2.32 \pm 0.05$ & 0.53 & 2.83  \\ \hline
d & $-1.17 \pm 0.01$ & $2.41 \pm 0.05$ & 0.55 & 2.82  \\ \hline
s & $-0.94 \pm 0.01$ & $5.83 \pm 0.08$ & 1.46 & 2.66  \\ \hline
c & $-0.81 \pm 0.01$ & $9.01 \pm 0.14$ & 3.5  & 3.41  \\ \hline
b & $-1.35 \pm 0.01$ & $7.16 \pm 0.07$ & 0.63 & 4.19  \\ \hline
\end{tabular}
$$
\begin{center}
{\bf Table II}\\
\end{center}

$$
\begin{tabular}{|c|c|}
\hline
\hline
\multicolumn{1}{|c|}{$P_{\perp}$} &
\multicolumn{1}{c|}{$R$} \\
\hline
3 &  0.55 \\ \hline
4 &  0.60 \\ \hline
5 &  0.64 \\ \hline
6 &  0.67 \\ \hline
7 &  0.67 \\ \hline
8 &  0.72 \\ \hline
9 &  0.72 \\ \hline
10 &  0.79 \\ \hline
11 &  0.80 \\ \hline
12 &  0.84 \\ \hline
13 &  0.86 \\ \hline
\end{tabular}
$$
\begin{center}
{\bf Table III}\\
\end{center}

\vfill\eject
\setlength{\textwidth}{16.cm}
\setlength{\textheight}{22.cm}

\end{document}